\pdfoutput=1
\documentclass[10pt,twocolumn,letterpaper]{article}
\usepackage[margin=0.75in,columnsep=0.28in]{geometry}
\usepackage{cite}
\usepackage{amsmath,amssymb,amsfonts}
\usepackage{algorithmic}
\usepackage{graphicx}
\usepackage{setspace}
\usepackage{algorithm,algorithmic}
\usepackage{hyperref}
\usepackage{booktabs}
\usepackage{multirow}
\usepackage{makecell}
\usepackage{adjustbox}
\usepackage{subcaption}
\hypersetup{hidelinks=true}
\usepackage{textcomp}
\usepackage{float}
\usepackage{tabularx}
\def\BibTeX{{\rm B\kern-.05em{\sc i\kern-.025em b}\kern-.08em
    T\kern-.1667em\lower.7ex\hbox{E}\kern-.125emX}}

\newcommand{\IEEEPARstart}[2]{#1#2}
\newenvironment{IEEEkeywords}%
  {\vspace{0.6\baselineskip}\noindent\textbf{Index Terms---}\ignorespaces}%
  {\par\vspace{0.6\baselineskip}}
\makeatletter
\newcommand\@titlenote[1]{%
  \begingroup\renewcommand\thefootnote{}\footnotetext{#1}\endgroup}
\long\def\thanks#1{\protected@xdef\@thanks{\@thanks
    \protect\@titlenote{#1}}}
\makeatother

\date{}

\begin{document}
\title{AsymFeX: A Symmetry-Driven Framework for Ischemic Stroke Segmentation Across Imaging Modalities and Stroke Stages}
\author{Maunil Shah, Vaanathi Sundaresan 
\thanks{This work was supported by DBT/Wellcome Trust India Alliance Fellowship [IA/E/22/1/506763]. This work was also supported in part by a grant from the Council of Scientific \& Industrial Research (CSIR) under its ASPIRE (Women Scientist Scheme) program [25WS(013)/2023-24/EMR-II/ASPIRE] and in part by Start-up Research Grant [SRG/2023/001406] from the Science and Engineering Research Board, India. VS is also supported by Pratiksha Trust, Bangalore, India [FG/PTCH-23-1004] and the Seed Research Grant [IE/RERE-22-0583] from the Indian Institute of Science, India.}
\thanks{Maunil Shah and Vaanathi Sundaresan are with the Department of Computational and Data Sciences, Indian Institute of Science, Bengaluru 560012, India (e-mail: maunilshah@iisc.ac.in, vaanathi@iisc.ac.in).}
}

\maketitle

\begin{abstract}
Fast and accurate segmentation of Acute Ischemic Stroke (AIS) lesions is essential for stroke prognosis and treatment planning. Non-contrast CT (NCCT), the first-line imaging modality for diagnosing ischemic infarcts, exhibits subtle infarct contrast, making manual delineation slow and labor-intensive. Motivated by this, and by the clinical practice of comparing brain hemispheres to localize infarcts, we propose a two-stage, nnU-Net-compatible 3D segmentation method. The first stage corrects head tilt to align each scan to its true anatomical mid-sagittal plane; the second applies a novel Asymmetric Feature Extraction (AsymFeX) module, comparing each voxel to its true contralateral counterpart within a local $3\times3\times3$ neighborhood via cross-hemispheric attention, feature disparity estimation, and dual-scale gating to capture both large and small infarcts. On AISD, our method achieves $0.6796$ Dice, $23.53$\,mm HD95, and $7.69$\,mL AVD, significantly outperforming existing state-of-the-art methods, with clinically relevant volumetric analysis at the 70\,mL thrombolysis-eligibility threshold. Proof-of-concept evaluation on ATLAS v2.1 and ISLES'24 demonstrates that the same symmetry-driven design generalizes across imaging modalities and stroke time points without architectural changes, further supported by an uncertainty analysis assessing reliability under clinical deployment. Code is publicly available at \url{https://github.com/biomedia-lab/AIS-detection}.
\end{abstract}


\section{Introduction}
\label{sec:introduction}
\IEEEPARstart{I}{schemic} stroke is a subtype of cerebrovascular stroke that occurs due to reduced cerebral blood flow caused by thrombotic or embolic events that block the cerebral vessels, leading to insufficient brain tissue perfusion \cite{musuka2015diagnosis}. According to the GBD 2019 report, there were 7.63 million cases of ischemic stroke, accounting for 62.4\% of all stroke incidents \cite{ding2022global}\cite{10.3389/fneur.2023.1309931}. During an ischemic event, brain tissue that incurs irreversible damage is called the infarct or ischemic core, while the tissue that is at risk of infarction but can potentially be salvaged is referred to as the ischemic penumbra.

Multiple imaging modalities are acquired depending on symptom onset time, patient criticality, and scanner availability, each guiding a distinct stage of reperfusion decision-making. NCCT is acquired to exclude intracranial hemorrhage or stroke mimics \cite{heit2015imaging}. Within the first 24 hours, Diffusion-Weighted Imaging (DWI) is the gold-standard study for detecting acute ischemia \cite{wintermark2013imaging}\cite{de2024isles}, owing to its higher sensitivity to infarction than NCCT \cite{https://doi.org/10.1111/ene.12849}. CT Perfusion (CTP) identifies salvageable, at-risk penumbra tissue, guiding whether a patient will benefit from reperfusion therapy \cite{heit2015imaging}\cite{https://doi.org/10.1111/ene.12849}. Together, NCCT, DWI, and CTP offer complementary information across the acute window to inform treatment decisions. In the chronic stage, beyond three weeks post-onset, T1-weighted MRI's high spatial resolution clearly reveals infarcts as hypointense regions \cite{bachtiar2024non}\cite{liew2022large}, supporting post-stroke treatment and rehabilitation planning. Given the time-sensitive nature of Acute Ischemic Stroke (AIS), delivering appropriate therapy as early as possible remains critical to salvaging maximal penumbra volume.

Although AIS infarcts are most appreciable on DWI in the acute phase, NCCT remains indispensable due to its shorter acquisition time, wide availability, and status as the first-line imaging modality in most stroke assessment protocols. However, AIS infarcts exhibit very subtle contrast on NCCT relative to healthy tissue \cite{heit2015imaging}\cite{wintermark2013imaging}, may suffer from artifacts and low signal-to-noise ratio, making it difficult to isolate affected tissue. Consequently, manual delineation of AIS lesions is laborious, time-consuming, and demands significant clinical expertise, motivating the need for automated segmentation methods that support rapid decision-making and treatment planning.

Deep learning (DL) has become the dominant paradigm for automated AIS segmentation, and a growing subset of recent work has attempted to exploit the brain's bilateral anatomical symmetry, since healthy hemispheres are approximately mirror-symmetric while an infarct locally distorts this symmetry. However, existing symmetry-aware methods remain limited in at least one of the following respects: they permit attention over the entire brain or entire contralateral hemisphere rather than a localized landmark-level comparison incurring unnecessary computational cost; they operate in 2D and cannot capture volumetric lesion characteristics; they compare a scan against its mirrored copy without first correcting head tilt, so the compared regions may not be true anatomical counterparts; or they require full brain slices as input, making them incompatible with the patch-based sampling used by modern segmentation frameworks such as nnU-Net. Beyond these architectural limitations, existing methods are typically evaluated on single imaging modality, single stroke time-point data and report no measure of prediction reliability. A detailed technical critique of representative methods is provided in Section~\ref{sec:related_work}.

Motivated by these gaps, we propose a novel framework: a two-stage, nnU-Net-compatible 3D segmentation pipeline that first geometrically aligns each scan to its true anatomical mid-sagittal plane, then exploits the resulting hemispheric correspondence through a localized, landmark-level cross-attention mechanism. Our contributions are summarized as follows: 
\begin{itemize}
    \item A 2-stage 3D segmentation pipeline that integrates with nnUNet and incorporates stroke-specific prior knowledge via an explicit geometric tilt-correction step, ensuring the contralateral comparison is anatomically valid.
    \item A novel Asymmetric Feature Extraction (AsymFeX) module combining 3D local cross-hemispheric attention, disparity estimation, and dual-scale gating, restricting comparison to a local neighborhood around each voxel's true contralateral landmark rather than the entire opposite hemisphere.
    \item A comprehensive evaluation against existing state-of-the-art and baseline methods on AISD, with extensive ablation isolating the contribution of each architectural component and volumetric analysis assessing clinical relevance.
    \item A proof-of-concept demonstration that the same asymmetry-driven design adapts across the acute-to-chronic stroke imaging continuum (CT perfusion and MRI).
    \item A comprehensive uncertainty quantification covering error-uncertainty analysis and calibration analysis on three datasets to assess prediction reliability across modalities and stroke stages.
\end{itemize}
The remainder of this paper is organized as follows. Section~\ref{sec:related_work} reviews related work and Section~\ref{sec:methodology} presents the proposed methodology. Section~\ref{sec:exp_setup} describes the datasets and experimental setup, and Section~\ref{sec:exps} reports results, including state-of-the-art comparison, computational cost, volumetric analysis, ablation studies, cross-modality generalization, and uncertainty quantification, followed by discussion and future directions in Section~\ref{sec:disc_conc}.

\section{Related Work}
\label{sec:related_work}
In this section, we present a literature survey pertaining to generic deep learning methods for fully supervised medical image segmentation, followed by methods specifically intended to segment AIS lesions by leveraging asymmetry.

Fully supervised DL for medical image segmentation has progressed markedly since the introduction of U-Net \cite{ronneberger2015u}, with subsequent architectures building on the encoder-decoder paradigm across both 2D and 3D settings. U-Net++ \cite{zhou2018unet++} introduced nested skip pathways to bridge the semantic gap between encoder and decoder features, improving delineation of fine anomalous regions. Attention U-Net \cite{oktay2018attention} incorporated attention gates to address low tissue contrast and shape/size heterogeneity across organs. The nnU-Net \cite{isensee2021nnu} instead emphasized systematic pipeline design, categorizing parameters such as voxel spacing, class proportions, learning rate, optimizer, post-processing, hardware constraints, etc. into fixed, rule-based, and empirical groups, yielding a task-agnostic framework that, despite relying on a standard U-Net backbone, remains a highly competitive segmentation baseline.

Transformer-based models quickly became a widely-adapted model for image segmentation due to their ability for effectively learning the long range dependencies. Swin-UNET \cite{cao2022swin} was the first pure transformer-based model proposed for medical image segmentation. However, hybrid models combining CNNs and transformers yields better performance as opposed to pure transformer-based models due to their ability of capturing complementary information. As such, Swin-UNETR \cite{hatamizadeh2021swin} is a hybrid variant where the encoder consists of Swin transformer block while the residual blocks and the decoder blocks comprise of convolution layers and deconvolution layers respectively. TransUNet \cite{chen2021transunet} proposed a hybrid encoder wherein the initial features are extracted using a CNN-based encoder followed by transformer encoder to complement the CNN features.

Recently, state-space architectures have emerged to address the computational complexity of transformer-based networks. U-Mamba \cite{ma2024u} integrates State Space Models (SSMs), specifically the Mamba selective scan mechanism, within a U-Net framework to model long-range global dependencies. By achieving linear computational complexity ($\mathcal{O}(N)$) relative to sequence length, U-Mamba overcomes the quadratic scaling limitations ($\mathcal{O}(N^2)$) inherent to Vision Transformers while maintaining high representative capacity.

Kolmogorov-Arnold Networks (KANs) benefit from interpretability and learnable activation functions \cite{liu2025kan}. Motivated by this, U-KAN \cite{li2025u} substitutes traditional Multi-Layer Perceptrons (MLPs) with tokenized KAN blocks near the network bottleneck. By leveraging KAN, the authors take a positive step in the direction of theory-backed medical image segmentation.

Methods discussed so far are general purpose medical image segmentation methods and do not utilize any stroke-specific information such as bilateral asymmetry. In this section, we discuss recent methods that have attempted to leverage the brain's anatomy-related cues and structural information to segment AIS lesions.
MDAN \cite{bao2021mdan} leverages hemispheric symmetry by learning representations that bring healthy tissue closer together while pushing lesion tissue apart, and integrates Mirror Feature Fusion modules along the skip connections to fuse original and mirrored features.
SAN-Net \cite{yu2023san} addresses cross-site MR heterogeneity through adaptive normalization and a symmetry-inspired augmentation strategy that focuses the model on a single hemisphere to simplify lesion localization. IS-Net \cite{yang2023net} aggregates features across multiple stages and employs a Nonlocal Parallel Decoder combining deformable convolution with self-attention to promote mask continuity and exploit hemispheric asymmetry.
PAPL \cite{sun2024pathological} proposes a three-stage framework that first learns representations of pathological asymmetry through contrastive pre-training, then carries this knowledge into end-to-end training, and finally refines mis-segmented regions using the predicted region and its counterpart across the mid-sagittal line. Cl-SegNet \cite{kuang2024hybrid} fuses transformer and CNN features efficiently and introduces a Bilateral Difference Learning (BDL) module to compare intensities between a feature map and its flipped copy.
More recently, Sun et al. \cite{sun2026deep} propose a two-stage method that first identifies suspicious lesion areas by learning to replicate a maximum intensity projection map, then uses this signal to predict probability maps for the original scan, the asymmetry map, and the final lesion mask. Li et al. \cite{li2025brain} propose DySym-UNet, which introduces a Symmetric Cross-Attention block to capture abnormalities between the left and right brain hemispheres.

Although existing methods have shown promising results, they carry limitations that constrain their use for 3D AIS segmentation. Several rely on attention over the entire brain (e.g., IS-Net) or entire contralateral hemisphere (e.g., DySym-UNet), which is computationally expensive and unnecessary, since detecting abnormality only requires comparison with the mirror-symmetric counterpart across the mid-sagittal line. Other methods (MDAN, PAPL, SAN-Net) are 2D and cannot capture 3D lesion characteristics such as surface texture or spatial continuity across slices. Cl-SegNet compares a feature map with its flipped copy without correcting head tilt beforehand, so misaligned scans yield a flipped copy that no longer corresponds to the true contralateral region, misleading the resulting feature difference. Further, SAN-Net, IS-Net, PAPL, and DySym-UNet all require full brain slices to exploit hemispheric disparity, making them incompatible with nnU-Net's random patch-based sampling, where an individual patch may not contain both hemispheres.

Beyond these architectural constraints, we identify two further gaps common across this body of work. First, existing methods are typically validated on a single imaging modality and at a single stroke time-point data, and do not demonstrate that a symmetry-driven design generalizes across the acute-to-chronic continuum spanning from NCCT to CT perfusion to MRI. Second, none report any measure of prediction reliability or uncertainty estimates, despite its importance for clinical deployment. 

The proposed method addresses gaps identified above using a robust 3D framework that integrates with nnU-Net and mimics a clinician's approach of comparing tissue directly with its contralateral landmark across the mid-sagittal line, while addressing the validation, generalization, and reliability. Section~\ref{sec:methodology} presents the proposed framework in detail.

\section{Methodology}
\label{sec:methodology}
\begin{figure*}[!ht]
\centering 
\includegraphics[width=0.80\textwidth]{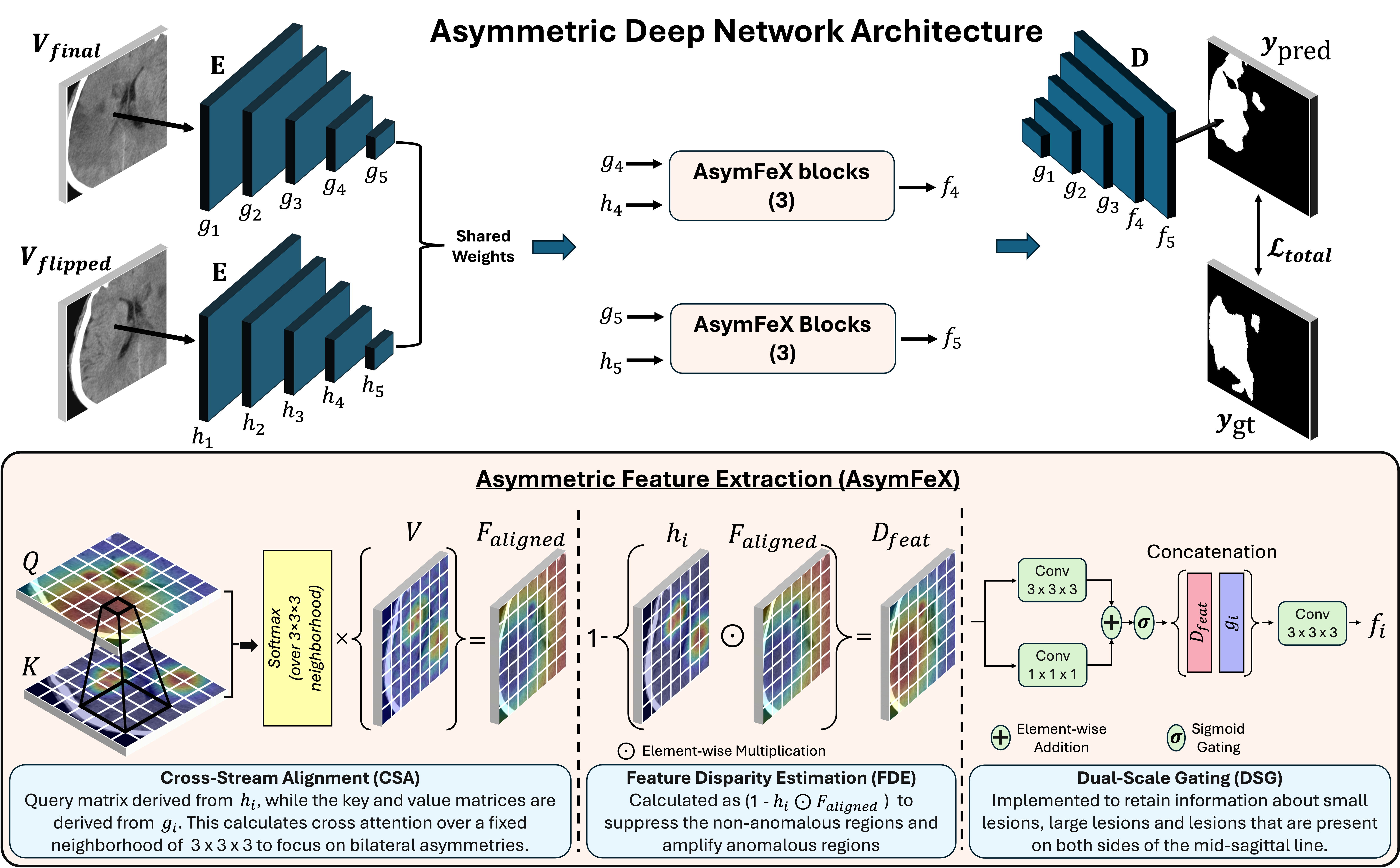} 
\caption{\textbf{Illustration of Stage 2 of the proposed methodology:} Stage 2 receives $V_{final}$ and $V_{flipped}$ from stage 1 as the input and uses AsymFex blocks along the deeper layers for leveraging the asymmetric differences. The AsymFex comprises of 3 sub-modules, viz., Cross-Stream Alignment (CSA), Feature Disparity Estimation (FDE) and Dual-Scale Gating (DSG).}
\label{fig:my_image} 
\end{figure*}

The proposed framework comprises two modality-agnostic stages, illustrated in Algorithm~\ref{alg:Geometric_Tilt_Correction} and Fig.~\ref{fig:my_image}. Stage 1 geometrically aligns the input volume to its true mid-sagittal plane, ensuring a subsequent vertical flip yields a valid contralateral counterpart. Stage 2 then processes the aligned volume and its flip through a shared-weight encoder and a novel Asymmetric Feature Extraction (AsymFeX) module, comparing each tissue region against its contralateral landmark to identify asymmetries indicative of infarction. Both stages are defined independently of any specific imaging modality, with modality-specific adaptation and the datasets are detailed in Sections~\ref{subsec:mod_spfc_adap} and~\ref{ssec:Datasets}, respectively.

\begin{algorithm}[t!] 
\caption{Geometric Tilt Correction.}
\label{alg:Geometric_Tilt_Correction}
\begin{algorithmic}[1]
\REQUIRE Ischemic CT Volume $V_{\text{raw}}$, Target Mask $M_{\text{raw}}$
\ENSURE Aligned and Cropped ($V_{\text{final}}$, $M_{\text{final}}$)
\STATE \textit{// 1. Optimal Slice Selection ($s_{\text{opt}}$)}
\STATE $V_{\text{stripped}} \leftarrow \text{FSL\_BET}(V_{\text{raw}})$
\STATE $V_{\text{bin}} \leftarrow \text{Binarize}(V_{\text{stripped}}, \tau)$ \COMMENT{Set $\ge \tau$ to 0 and $>0$ to 1}
\STATE $\text{Best\_Score} \leftarrow -1, \quad s_{\text{opt}} \leftarrow \text{None}$

\FOR{each slice $i$ in $V_{\text{bin}}$}
    \STATE $S_i \leftarrow V_{\text{bin}}[:, :, i]$
    \STATE $S_{\text{filled}} \leftarrow \text{Fill\_Holes\_2D}(S_i)$
    \STATE $L_i \leftarrow \text{Label\_Connected\_Components}(S_{\text{filled}})$
    \STATE $\tilde{S}_i \leftarrow \text{Isolate\_Largest\_Component}(L_i)$
    \STATE $\text{Score} \leftarrow \sum \tilde{S}_i$ \COMMENT{Calculate component area}
    \IF{$\text{Score} > \text{Best\_Score}$}
        \STATE $\text{Best\_Score} \leftarrow \text{Score}$
        \STATE $s_{\text{opt}} \leftarrow V_{\text{stripped}}[:, :, i]$
    \ENDIF
\ENDFOR
\STATE \textit{// 2. Symmetry-Driven Tilt Correction}
\STATE $C(C_x, C_y) \leftarrow \text{Compute\_Centroid}(s_{\text{opt}})$
\STATE $V_{\text{trans}} \leftarrow \text{Translate\_Volume}(V_{\text{stripped}}, C)$
\STATE $M_{\text{trans}} \leftarrow \text{Translate\_Volume}(M_{\text{raw}}, C)$

\STATE $\text{Min\_Disparity} \leftarrow \infty, \quad \theta \leftarrow 0.0$
\FOR{$\phi = -45.0^\circ \text{ to } 45.0^\circ \text{ with step } 0.05^\circ$}
    \STATE $s_{\text{rot}} \leftarrow \text{Rotate\_Slice}(s_{\text{opt}}, R_\phi)$
    \STATE $H_{\text{left}}, H_{\text{right}} \leftarrow \text{Split\_Vertical\_Axis}(s_{\text{rot}})$
    \STATE $H_{\text{flipped\_right}} \leftarrow \text{Horizontal\_Flip}(H_{\text{right}})$
    \STATE $\text{Disparity} \leftarrow |H_{\text{left}} \cup H_{\text{flipped\_right}}| - |H_{\text{left}} \cap H_{\text{flipped\_right}}|$
    \IF{$\text{Disparity} < \text{Min\_Disparity}$}
        \STATE $\text{Min\_Disparity} \leftarrow \text{Disparity}$
        \STATE $\theta \leftarrow \phi$
    \ENDIF
\ENDFOR

\STATE $V_{\text{aligned}} \leftarrow \text{Rotate\_Volume}(V_{\text{trans}}, \theta)$
\STATE $M_{\text{aligned}} \leftarrow \text{Rotate\_Volume}(M_{\text{trans}}, \theta)$
\STATE \textit{// 3. Tight Cropping}
\STATE $V_{\text{final}}, M_{\text{final}} \leftarrow \text{Tight\_Bound\_Box\_Crop}(V_{\text{aligned}}, M_{\text{aligned}})$
\RETURN $V_{\text{final}}, M_{\text{final}}$
\end{algorithmic}
\end{algorithm}

\subsection{Stage 1: Geometric Tilt Correction}
Most existing NCCT datasets \cite{liang2021symmetry} are not tilt corrected. To truly leverage symmetric differences in the downstream asymmetric learning task, we first implement a pre-processing algorithm that rectifies head tilt in a CT scan and its corresponding segmentation mask. The complete workflow is illustrated in Algorithm \ref{alg:Geometric_Tilt_Correction}.

\subsubsection{Optimal Slice Selection}
We begin with applying a brain windowing (width = 80, level = 40) and mapping the voxel values to the range [0, 255]. We then proceed to remove the non-brain structures, such as the CT scanner head cushion, using the FSL Brain Extraction Tool \cite{smith2002fast}. Following CT cushion removal, we select an optimal axial slice that captures the maximum cross-sectional area of the brain parenchyma.

To isolate the brain structure from high-intensity skull elements, the stripped volume $V_{\text{stripped}}$ is binarized using an intensity threshold $\tau = 250$ to reliably separate residual high-intensity skull fragments from brain parenchyma following skull-stripping. Voxels with attenuation values greater than or equal to $\tau$ are suppressed to $0$, while all the remaining voxel values are mapped to $1$.

For each axial slice $i$ along the longitudinal axis of the binarized volume, a 2D morphological hole-filling operation is executed to create a solid, continuous representation of the brain structure. We then apply 2D connected component labeling using a $3\times3$ structural connectivity matrix to eliminate peripheral artifacts and isolate the largest contiguous connected component mask $\tilde{S}_i$. The structural score for the slice $i$ is defined as the total area (voxel sum) of this refined tissue region:
\begin{equation}
\text{Score}(i) = \sum_{x,y} \tilde{S}_i(x,y)
\end{equation}
The slice index exhibiting the maximum area score is designated as the optimal slice $s_{\text{opt}}$
\begin{equation}
s_{\text{opt}} = \arg\max_{i} \text{Score}(i)
\end{equation}

\subsubsection{Symmetry-Driven Tilt Correction}
Once $s_{\text{opt}}$ is identified, its geometric centroid $C = (C_x, C_y)$ is computed based on the spatial coordinates of its non-zero pixels. The entire 3D volume ($V_{\text{stripped}}$) and the segmentation mask ($M_{\text{raw}}$) are then translated to $C$. 

Using a grid search brute-force optimization paradigm, the exact head tilt angle $\theta$ is extracted from the translated optimal slice ($s_{\text{opt}}$) by minimizing hemispheric structural disparity. This is achieved by defining a search space for angle in the range $\phi \in (-45^\circ, 45^\circ)$ at increments of $0.05^\circ$. For each candidate angle $\phi$:
\begin{enumerate}
    \item A standard 2D rotation matrix $R_\phi$ is constructed, and $s_{\text{opt}}$ is rotated about its centered origin.
    \item The rotated slice is partitioned along its vertical line into left and right halves.
    \item The right half is flipped horizontally and overlaid onto the left half.
    \item A hemispheric disparity score, defined as the difference between the union and the intersection of the two halves, is calculated.
\end{enumerate}
The optimal tilt correction angle $\theta$ is determined as the angle that minimizes this asymmetry disparity score, effectively mapping the precise true vertical symmetry line of the cranium. Mathematically, this optimization is expressed as:
\begin{equation}
\theta = \arg\min_{\phi} \left( |H_{\text{left}} \cup H_{\text{flipped\_right}}| - |H_{\text{left}} \cap H_{\text{flipped\_right}}| \right)
\end{equation}

\subsubsection{Tight Cropping}
Following calibration, the translated 3D CT volume ($V_{\text{trans}}$) and its corresponding ground-truth segmentation mask ($M_{\text{trans}}$) are rotated by the optimal angle $\theta$, yielding pairs $V_{\text{aligned}}$ and $M_{\text{aligned}}$. To eliminate redundant background voxels and constrain the region of interest strictly to the cranial anatomy, a tight 3D bounding box crop is applied, producing the finalized volume $V_{\text{final}}$ and mask $M_{\text{final}}$. We then mirror the volume $V_{\text{final}}$ to produce $V_{\text{flipped}}$ which swaps the left and right hemisphere so that each landmark becomes aligned with its counterpart on the opposite side. This flipped volume ($V_{\text{flipped}}$) is then coupled with $V_{\text{final}}$ to create a dual-stream input.

\subsection{Stage 2: Asymmetric Deep Network Architecture}
The fundamental objective of the proposed architecture is to exploit asymmetry between the two bilateral hemispheres by leveraging the tilt-corrected volume $V_{\text{final}}$ along with its horizontally flipped version $V_{\text{flipped}}$ for direct comparison. The overall architecture is depicted in Fig. \ref{fig:my_image}.

\subsubsection{Shared CNN Encoder}
\label{sssection:shared_CNN_enc}
Inspired by the architectural topology of the 3D U-Net \cite{ronneberger2015u}, the framework first passes the inputs through a five-stage convolutional encoder, denoted as $\mathbf{E}$. Each stage within $\mathbf{E}$ contains two consecutive blocks, where each block is composed of a standard 3D convolutional layer followed by a 3D instance normalization layer and a Rectified Linear Unit (ReLU) activation function. The two input streams, the tilt-corrected volume $V_{\text{final}} \in \mathbb{R}^{1 \times D \times H \times W}$ and its mirrored contralateral counterpart $V_{\text{flipped}} \in \mathbb{R}^{1 \times D \times H \times W}$, are processed independently through this encoder. The encoder $\mathbf{E}$ operates under a weight-sharing configuration across both streams. The choice of a pure CNN architecture for the encoder is driven by three reasons. Firstly, it extracts similar features from both inputs and maps them into a shared space. Since both hemispheres are processed using the exact same weights, any differences found between their feature maps later in the network could be attributed to actual tissue differences, which could be further leveraged by the \textit{Asymmetric Feature Extraction (AsymFeX)} modules (\ref{sssec:symfex_method}). Secondly, our model is integrated within the nnU-Net \cite{isensee2021nnu} framework, which relies on random 3D patch-based sampling during training. Because an isolated 3D patch rarely captures the entire brain, the CNN encoder ensures that the network establishes a clear representation of a voxel's local surroundings before the model attempts to evaluate broader hemispheric differences. Thirdly, this setup closely mimics clinical workflow where a radiologist typically inspects one hemisphere first to establish a baseline of normal anatomy before cross-referencing the opposite side to detect anomalies. Our encoder similarly extracts and preserves structural features independently before any comparison occurs.

Formally, let $\mathbf{E}(V_{\text{final}}) = \{g_1, g_2, g_3, g_4, g_5\}$ and $\mathbf{E}(V_{\text{flipped}}) = \{h_1, h_2, h_3, h_4, h_5\}$ denote the multi-scale feature maps extracted by $\mathbf{E}$ from $V_{\text{final}}$ and $V_{\text{flipped}}$, respectively. We then route the deep features from the last two layers, namely $(g_4, h_4)$ and $(g_5, h_5)$, forward into the corresponding Asymmetric Feature Extraction (AsymFeX) blocks along the skip connections. We place AsymFeX only in deeper stages, where features are semantically richer and spatially broader, allowing it to capture meaningful anatomical differences rather than shallow texture or edge information.

\subsubsection{Asymmetric Feature Extraction (AsymFeX)}
\label{sssec:symfex_method}
Building on the independent features extracted by the shared encoder $\mathbf{E}$, the asymmetric comparison modules are integrated at feature levels 4 and 5, which contain 256 and 320 channels, respectively. This module consists of a cascaded sequence of identical AsymFeX blocks. Within each block, three stages are performed sequentially: Cross-Stream Alignment (CSA), Feature Disparity Estimation (FDE), and Dual-Scale Gating (DSG).

\textbf{(a) Cross-Stream Alignment (CSA)}: 
\label{heading:csa}
Standard global self-attention scales quadratically with the number of voxels ($\mathcal{O}(N^2)$) \cite{vaswani2017attention}, making it expensive for 3D image analysis. Moreover, it allows every voxel to attend to distant regions, which is unnecessary for our objective as we only need to compare each tissue region with its corresponding contralateral counterpart in the flipped input. Since our Stage~1 pre-processing ensures that corresponding contralateral features are nearly aligned, we restrict cross-attention between $g_i$ and $h_i$ to a local $3\times3\times3$ spatial neighborhood (found to be an optimal choice - refer to Section~\ref{sssec:neighborhood} for further experimental details), giving a neighborhood size of $N_w = 27$ voxels. This reduces the attention cost from $\mathcal{O}(N^2)$ to $\mathcal{O}(N \cdot N_w)$ with $N_w \ll N$, i.e. linear in the number of voxels.

The local neighborhood is motivated by the fact that even after geometric alignment, the two hemispheres may not match perfectly at the voxel level due to the brain's natural anatomical asymmetry, interpolation effects and minor tilt-correction errors. Therefore, a true contralateral match may not lie exactly at the mirrored location, so the window provides a search region around the expected landmark which is large enough to absorb these errors (ablation conducted with different window sizes as specified in Section~\ref{sssec:neighborhood}).

We denote a feature volume by its shape $(C, H, W, D)$, where $C$ is the number of channels and $(H, W, D)$ are the height, width, and depth of the 3D spatial grid. We start by computing the Query ($Q$), Key ($K$) and Value ($V$) tensors. The $Q$ is projected from the flipped stream ($h_{\text{i}}$) while the $K$ and $V$ are projected from the original stream ($g_{\text{i}}$), each using a separate point-wise $1\times1\times1$ convolution preserving the tensor shape $(C,H,W,D)$. This can be denoted as:
\begin{subequations}
\begin{align}
Q &= W_q(h_{\text{i}}), \\
K &= W_k(g_{\text{i}}), \\
V &= W_v(g_{\text{i}})
\end{align}
\end{subequations}
Next, the $C$ channels are partitioned into $M$ attention heads, each operating on $d_k = C/M$ channels, so that similarity is measured independently within $M$ feature subspaces. Hence, the matrices $Q, K, V$ are reshaped to $\mathbb{R}^{M\times H\times W\times D\times d_k}$. The local attention window spans $3$ positions per axis and therefore $N_w$ accumulates to $27$ neighbors per voxel. 
Next, before gathering neighbors in $K$ and $V$, we first zero-pad them by one voxel on each spatial side to include the boundary voxels as well. We then apply three successive tensor unfold operations, one along each spatial axis ($H, W, D$), which collects the $3\times3\times3$ block of neighbors for every voxel. More precisely, for each voxel we collect the neighbors obtained by applying the offsets $(dx, dy, dz) \in \{-1, 0, 1\}^3$ along the three spatial axes, i.e. the voxel itself together with its immediate neighbors in every direction. Upon flattening the three window axes yields,
\begin{equation}
\tilde{K},\,\tilde{V} \in \mathbb{R}^{M\times H\times W\times D\times N_{w}\times d_k}
\end{equation}
In other words, $\tilde{K}$ and $\tilde{V}$ make each voxel's $N_w=27$ neighbors directly available along a dedicated axis, so that the subsequent attention reduces to vectorized multiplications. For head $m$, anchor voxel $(x,y,z)$, and neighbor index $n \in \{1, 2, .., N_{w}\}$, the unnormalized score $U_m$ is calculated as the product between the $Q$ and the gathered neighbor key $\tilde K$, taken over the $d_k$ channels of the head:
\begin{equation}
U_m(x,y,z,n) = \sum_{c=1}^{d_k}
Q(m,x,y,z,c)\;
\tilde K(m, x,y,z,n,c) \label{eq:scores}
\end{equation}
Because the zero-padding introduces neighbors that lie outside the true feature grid for voxels near the boundary, an additive mask $M$ removes their influence:
\begin{equation}
M(x,y,z,n)=
\begin{cases}
0 & \text{if } (x+dx,\,y+dy,\,z+dz)\in\mathcal{B},\\[2pt]
C_{\text{mask}} & \text{otherwise}
\end{cases}
\end{equation}
where $\mathcal{B}$ is the set of in-bounds voxel positions of the feature grid and $C_{\text{mask}}$ is a large negative constant ($-10^4$). Next, the scores obtained in Eqn.~\ref{eq:scores} are scaled by a factor of $1 /\sqrt{d_{k}}$, augmented with the boundary mask $M$, and normalized by a softmax over the $N_w$ neighbors:
\begin{align}
S_m(x,y,z,n) &= \frac{1}{\sqrt{d_k}}\,U_m(x,y,z,n) + M(x,y,z,n) \\
A_m(x,y,z,n) &= \operatorname{Softmax}_{\,n}\!\big(S_m(x,y,z,n)\big)
\end{align}
where $A_m(x,y,z,n)$ represents attention weight of each neighbor for the voxel $(x, y, z)$ under the head $m$. The final attention for head $m$ ($Attn_m \in \mathbb{R}^{D\times H\times W\times d_k}$) is calculated as: 
\begin{equation}
Attn_m(x, y, z, d_k) = \sum_{n=1}^{N_w} A_m(x,y,z,n)\;
\tilde{V}(m,x,y,z,n,d_k)
\
\end{equation}
Likewise, each head produces its output as the attention-weighted sum of the gathered neighbor values. The per-head outputs are concatenated and passed through a final point-wise projection $W_o$ to form the aligned feature volume:
\begin{align}
F_{\text{aligned}} &= W_o\!\left(\operatorname*{\big\Vert}_{m=1}^{M}
Attn_m \right)
\end{align}
where $\big\Vert$ denotes concatenation across heads. Hence, the CSA module helps the model to isolate attention to a local neighborhood of $27$ voxels thereby allowing it to focus only on relevant bilaterally opposite landmarks.

\textbf{(b) Feature disparity estimation (FDE)}: Once the CSA module captures the neighborhood information, we need a way to guide the model into learning asymmetric features. Hence, we calculate the feature disparity between $F_{\text{aligned}}$ and the encoder output for the flipped input $h_i$, where we first apply an $L_2$ vector normalization across the channel dimension for both the locally aligned feature map $F_{\text{aligned}}$ and the flipped contralateral feature map $h_{i}$. The Feature Disparity map ($D_{\text{feat}}$) is then calculated as
\begin{equation}
D_{\text{feat}} = 1.0 - \left( \frac{F_{\text{aligned}}}{\|F_{\text{aligned}}\|_2 + \epsilon} \odot \frac{h_i}{\|h_i\|_2 + \epsilon} \right)
\end{equation}
where $\odot$ denotes element-wise multiplication and $\epsilon = 10^{-6}$ is added to prevent division-by-zero. The factor $1-(\cdot)$ yields a disparity map. The outcome $D_{\text{feat}}$ exhibits higher values where the aligned and the contralateral features diverge.  

\textbf{(c) Dual-scale gating (DSG)}: Ischemic lesions can vary widely in size, ranging from large MCA (Middle Cerebral Artery) obstructions to tiny, isolated focal infarcts spanning only a few voxels. Hence, it is crucial to retain the information about tiny lesions in the deeper features. 

To address this, the model processes the feature disparity map ($D_{\text{feat}}$) through a parallel dual-scale gating module designed to capture both broad regional trends and sharp focal details. As shown in Fig. \ref{fig:my_image}, the disparity map is split into two paths:
\begin{enumerate}
    \item Gate Small (GS): Employs a convolution block with a kernel size of $1 \times 1 \times 1$. Since tiny infarcts span only a few voxels, a larger kernel dilutes their signal with surrounding information; the $1 \times 1 \times 1$  kernel instead operates on each voxel independently, preserving these faint signals.
    \begin{equation}
    P_{\text{small}} = \text{Conv}_{1\times1\times1}(D_{\text{feat}})
    \end{equation}
    
    \item Gate Large (GL): Uses a standard $3 \times 3 \times 3$ convolution. Larger lesions span many voxels, so a $3 \times 3 \times 3$ kernel aggregates local context to capture their extent, while avoiding bigger kernels whose cubically-growing parameter cost tends to blur lesion boundaries.
    \begin{equation}
    P_{\text{large}} = \text{Conv}_{3\times3\times3}(D_{\text{feat}})
    \end{equation}
\end{enumerate}

The outputs from both paths are recombined via element-wise addition. We then apply Group Normalization ($\text{GN}$) followed by a Sigmoid activation function ($\sigma$) to generate a spatial gating map $G \in [0,1]$:
\begin{subequations}
\begin{align}
G = \sigma\left(\text{GN}\left(P_{\text{large}} + P_{\text{small}}\right)\right) \\
G_{\text{aligned}} = G \odot F_{\text{aligned}}
\end{align}
\end{subequations}

Next, the representation $G_{\text{aligned}}$ is concatenated with the original unaltered feature map $g_{\text{i}}$. This safeguards rare cases where infarcts appear at bilaterally opposite points. In such cases, since both landmarks look similar, their disparity is low and the gating would suppress the signal. Including $g_{i}$ preserves information about such lesions. We then apply a convolution to fuse the concatenated features and project the $2C$ channels back to $C$, followed by Group Normalization (GN) and Gaussian Error Linear Unit (GELU) activation function as follows:
\begin{subequations}
\begin{gather}
f_{\text{i}} = \text{GELU}\left(\text{GN}\left(\text{Conv}_{3\times3\times3}\left([G_{\text{aligned}} \,\, \Vert \,\, g_{\text{i}}]\right)\right)\right)
\end{gather}
\end{subequations}
where $\Vert$ denotes tensor concatenation along the channel axis.

\subsubsection{CNN Decoder}
The features $(g_1, g_2, g_3, f_4, f_5)$ are passed to the decoder $\mathbf{D}$ to construct the final segmentation map. The flipped stream $h_i$ serves only as a contralateral reference for computing inter-hemispheric disparity. Once this comparison is complete, the segmentation is produced in the original, unflipped anatomical space. The decoder progressively upsamples the feature maps using 3D transposed convolutions, recombining them with the corresponding encoder features via skip connections. The output from the final layer contains $N_{cls}+1$ channels, $N_{cls}$ being the number of foreground classes, followed by a softmax activation function to produce per-voxel class probabilities. At inference, class predictions are then obtained by taking the argmax of the probabilities.

\subsubsection{Loss function}
During training, the loss function ($\mathcal{L}_{\text{total}}$) used is the sum of Dice loss ($\mathcal{L}_{\text{Dice}}$) and Cross-Entropy loss ($\mathcal{L}_{\text{CE}}$). This loss is implemented at all decoder layers to preserve stroke representations across all resolutions and can be represented as follows:
\begin{equation}
\mathcal{L}_{\text{total}} =
\sum_{i=1}^{5} \alpha_i
\left[
\mathcal{L}^{(i)}_{\text{CE}}\!\left(y_{\text{pred}}, y_{\text{gt}}\right)
+
\mathcal{L}^{(i)}_{\text{Dice}}\!\left(y_{\text{pred}}, y_{\text{gt}}\right)
\right]
\end{equation}
where $\alpha_i$ represents the weight of the loss function at the $i^{th}$ layer of the decoder $\mathbf{D}$.

\section{Experimental Setup}
\label{sec:exp_setup}
\subsection{Datasets} 
\label{ssec:Datasets}
We evaluate the proposed method on three publicly available datasets spanning the acute-to-chronic stroke imaging continuum.

\subsubsection{Acute Ischemic Stroke Dataset (AISD)}
\label{sssec:aisd}
The AISD \cite{liang2021symmetry} dataset comprises 397 NCCT scans acquired within 24 hours of symptom onset, with lesions manually delineated using DWI scans (acquired within 24 hours post-CT) as reference and were further reviewed by a senior doctor. In this study, we use only the NCCT scans for training and validation. We use the standard split of 345 training and 52 test scans.

\subsubsection{Anatomical Tracings of Lesions After Stroke (ATLAS v2.1) dataset}
\label{sssec:atlas}
ATLAS v2.1 \cite{liew2022large} is a multi-site, T1-weighted MRI dataset of chronic and sub-acute stroke, comprising 1271 volumes from 44 research cohorts. The data are derived from studies that were approved by their local ethics committee and were conducted in accordance with the 1964 Declaration of Helsinki. Informed consent was obtained from all subjects. The ethics committee at the receiving site (the University of Southern California) approved the receipt and sharing of the 
de-identifed data, which do not contain any personal identifers. We use the 655 publicly available training volumes with their masks, and evaluate on the 300 test volumes for which v2.1 additionally released ground-truth masks that were withheld in v2.0.

\subsubsection{Ischemic Stroke Lesion Segmentation Challenge (ISLES'24) dataset}
\label{sssec:isles}
ISLES'24 \cite{de2024isles,riedel2024isles} is a multi-center longitudinal dataset providing a pre-interventional CT trilogy (NCCT, CT-Perfusion, CT-Angiography), with lesion masks originally derived from follow-up DWI acquired 2-9 days later. This multi-center study used data from studies approved by their local ethics committees and were conducted in accordance with 1964 Declaration of Helsinki and its updated version 24. Due to defacing and rigorous anonymization, the ethics committee at the receiving site (University of Zurich) approved sharing the de-identified data. ISLES'24 targets prediction of post-interventional infarct from pre-interventional CT, but since infarcts evolve after the initial scan, the released masks do not reflect the pre-interventional state. An expert neuroradiologist therefore manually re-delineated pre-interventional ischemic core and penumbra on the CT perfusion scans using ITK-SNAP \cite{py06nimg}. Given a mean onset-to-door time of 3h:28m \cite{riedel2024isles} and the limited sensitivity of NCCT to early ischemic change at this stage \cite{heit2015imaging} \cite{wintermark2013imaging}, we incorporate CT perfusion parameter maps, including Cerebral Blood Volume (CBV), Cerebral Blood Flow (CBF), Mean Transit Time (MTT) and Time-to-Maximum ($T_{\max}$), alongside NCCT to improve core and penumbra visibility. As the official 96-case test split remains withheld, we partition the available 149 cases into 109 training and 40 test cases.

\subsection{Implementation details}
\label{ssec:imp_details}
All experiments were implemented in PyTorch 2.5.1 (CUDA 12.1) within the nnU-Net \cite{isensee2021nnu} codebase, using two NVIDIA RTX A6000 GPUs (48\,GB each). Models were trained for 160 epochs using Stochastic Gradient Descent (SGD) optimizer with a polynomial learning-rate schedule (initial rate $1\times10^{-3}$, exponent $0.9$) and batch size 6. Data augmentations include Gaussian noise ($\mu$ = 0, $\sigma \in [0.1, 0.3]$), Gaussian blur ($\mu$ = 0, $\sigma \in [0.5, 2.5]$), gamma transform ($\gamma \in [0.7, 1.5]$), linear contrast transform ($m \in [0.75, 0.125]$). 

\subsection{Evaluation metrics and statistical testing}
\label{ssec:eval_stat}
Segmentation performance is quantified using three complementary metrics: the Dice similarity coefficient measures volumetric overlap between the predicted mask and ground truth, the 95th-percentile Hausdorff Distance (HD95) (mm) quantifies boundary agreement while suppressing sensitivity to small outlier regions, computed as the 95th percentile of the set of nearest-neighbor distances between predicted and ground-truth surface points in both directions and lastly Absolute Volume Difference (AVD) (ml) captures the difference between the total predicted and the ground truth volume. 
To assess the statistical significance of the improvement in performance, we perform a paired Wilcoxon signed-rank test on a per-case basis for each metric and each dataset, with a significance threshold of $p < 0.05$. 

\section{Experiments and results}
\label{sec:exps}
\subsection{Comparison with existing state-of-the-art methods}
\label{subsec:SOTA_comp}
\begin{figure*}[!ht]
\centering 
\includegraphics[width=0.80\textwidth]{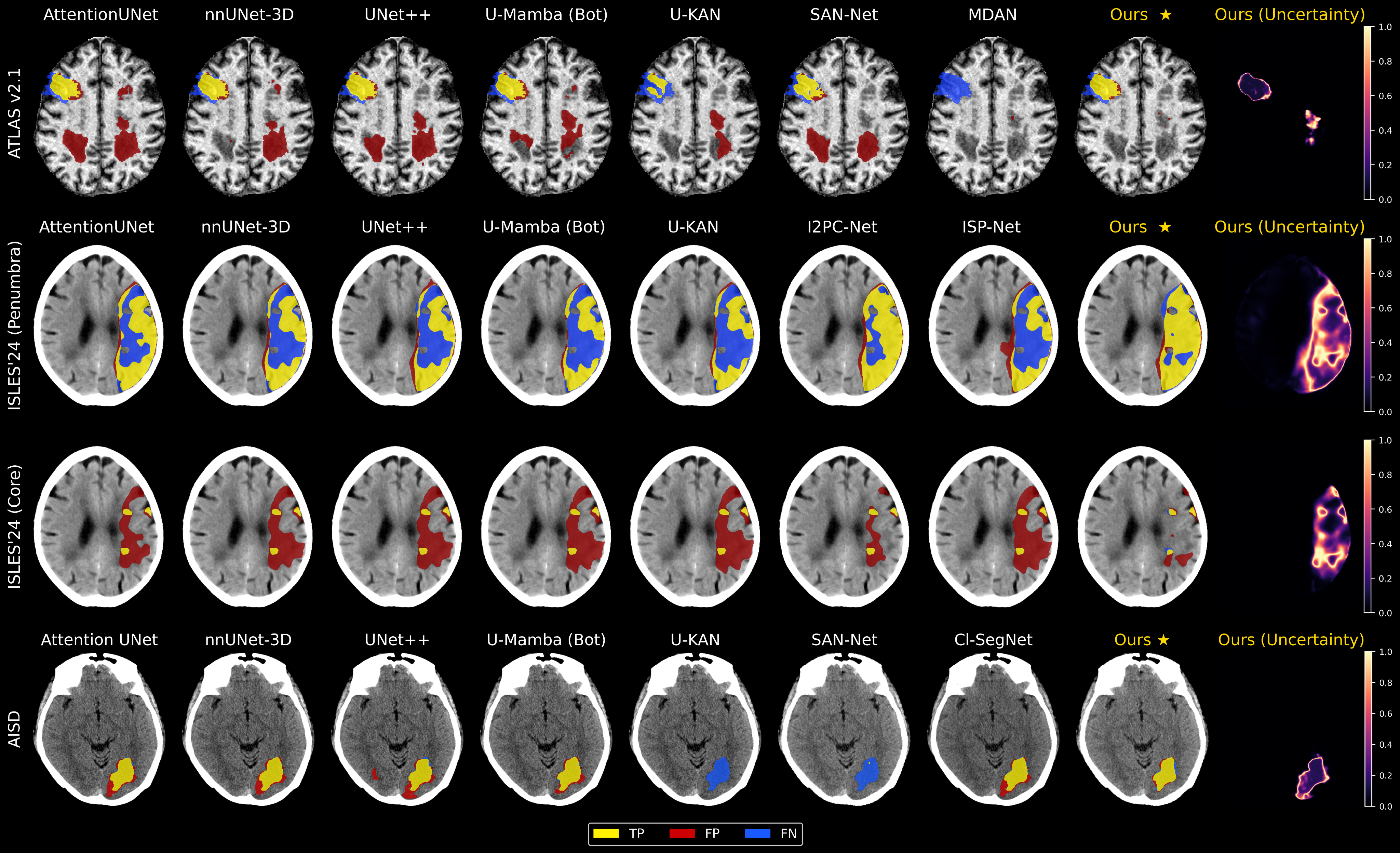} 
\caption{\textbf{Qualitative comparison of segmentation outputs:} Visual results of comparison of our method with existing methods along with Uncertainty maps on ATLASv2.1 (row 1), ISLES'24 (row 2 - Penumbra, row 3 - Core) and AISD (row 4).}
\label{fig:sota_comp} 
\end{figure*}

On the AISD dataset, we compare against baselines such as nnU-Net \cite{isensee2021nnu}, Attention U-Net \cite{oktay2018attention}, U-Net++ \cite{zhou2018unet++}, TransUNet \cite{chen2021transunet}, Swin-UNETR \cite{cao2022swin}, U-Mamba \cite{ma2024u}, and U-KAN \cite{li2025u}, all trained within the same nnU-Net codebase for fair comparison. Additionally, existing state-of-the-art methods are evaluated using their publicly available implementations (e.g., SAN-Net \cite{yu2023san}, Cl-SegNet \cite{kuang2024hybrid}) or by replicating methodology in their publications when code is unavailable (MDAN \cite{bao2021mdan}). For methods DySym-UNet \cite{li2025brain}) and PAPL \cite{sun2024pathological} we draw indirect comparison using their reported numbers as the AISD test split is standard and shared across all methods. Table~\ref{tab:seg_results} reports the results of the comparative analysis. The proposed method achieves a Dice of 0.6796, HD95 of 23.53, AVD of 7.69 and significantly outperforms all compared baselines and stroke-specific methods across Dice, HD95, and AVD ($p<0.05$, Wilcoxon signed-rank test). Our method also outperforms PAPL \cite{sun2024pathological} (Dice 0.5468) and DySym-UNet \cite{li2025brain} (Dice 0.6183) by 24.28\% and 9.91\% respectively whereas U-Mamba (Enc) variant fares the lowest (Dice 0.3451). 
Fig.~\ref{fig:sota_comp} shows a visual comparison against existing state-of-the-art approaches. U-KAN and SAN-Net miss the lesion entirely in the shown cases, while our method locates it reliably and traces the boundary more accurately than nnU-Net (3D), U-Net++, Attention U-Net, U-Mamba (Bot), and Cl-SegNet. These visual results align with Table~\ref{tab:seg_results}: the improvement in HD95 reflects the sharper boundaries, while the lower AVD indicates more accurate lesion-volume estimation, clinically important for treatment-eligibility decisions. Our method also has the lowest Dice standard deviation, indicating more consistent performance with fewer critical failures than U-KAN and SAN-Net.

\begin{table*}[t]
\centering
\caption{Quantitative comparison of the proposed method against existing
approaches on AISD and small lesions of AISD using 70mL as cutoff value. Best result in each
column is highlighted in \textbf{bold}. $\uparrow$: higher is better;
$\downarrow$: lower is better. * denotes statistically significant
difference in performance with p-value $<$ 0.05 using Wilcoxon signed rank test.}
\begin{adjustbox}{max width=\textwidth}
\begin{tabular}{l|c|c|c|c|c|c}
\hline
\multirow{2}{*}{\textbf{Method}} & \multicolumn{3}{c|}{\textbf{AISD (Overall)}} & \multicolumn{3}{c}{\textbf{AISD - Small Lesions (volume $\le$ 70mL)}} \\
\cline{2-7}
 & \makecell{\textbf{Dice} $\uparrow$ \\ (mean $\pm$ std.)} & \makecell{\textbf{HD95 (mm)} $\downarrow$ \\ (mean $\pm$ std.)} & \makecell{\textbf{AVD (mL)} $\downarrow$ \\ (mean $\pm$ std.)} & \makecell{\textbf{Dice} $\uparrow$ \\ (mean $\pm$ std.)} & \makecell{\textbf{HD95 (mm)} $\downarrow$ \\ (mean $\pm$ std.)} & \makecell{\textbf{AVD (mL)} $\downarrow$ \\ (mean $\pm$ std.)} \\
\hline
nnU-Net (3D) \cite{isensee2021nnu} & 0.5765 ± 0.2248 & 49.9350 ± 27.5323 & 12.0055 ± 16.9135 & 0.4963 ± 0.2672 & 52.9767 ± 28.9305 & 9.7279 ± 13.0793 \\
nnU-Net (2D) \cite{isensee2021nnu} & 0.5391 ± 0.2353 & 30.4997 ± 22.0304 & 11.4344 ± 13.9028 & 0.4685 ± 0.2519 & 34.2819 ± 23.8421 & 9.2140 ± 11.5344 \\
Attention U-Net \cite{oktay2018attention} & 0.5910 ± 0.2083 & 38.1049 ± 28.0969 & 13.4658 ± 24.3429 & 0.5127 ± 0.2629 & 41.1659 ± 30.2004 & 9.0711 ± 13.1302 \\
U-Net++ \cite{zhou2018unet++} & 0.5519 ± 0.2293 & 43.0069 ± 26.1526 & 14.8845 ± 24.7511 & 0.4620 ± 0.2540 & 46.0437 ± 26.0457 & 11.2365 ± 13.3084 \\
TransUNet \cite{chen2021transunet} & 0.3577 ± 0.2346 & 56.0839 ± 16.1364 & 22.3262 ± 26.4055 & 0.2840 ± 0.2160 & 57.7683 ± 16.4336 & 18.7202 ± 17.1059 \\
SwinUNeTR \cite{hatamizadeh2021swin} & 0.4813 ± 0.2255 & 49.3992 ± 23.2987 & 17.3289 ± 30.6559 & 0.4104 ± 0.2515 & 51.5550 ± 23.0180 & 10.6878 ± 14.4025 \\
U-Mamba Bot \cite{ma2024u} & 0.5473 ± 0.2295 & 45.0523 ± 23.0681 & 14.0949 ± 19.3919 & 0.4739 ± 0.2599 & 47.3243 ± 22.8482 & 10.7338 ± 10.4746 \\
U-Mamba Enc \cite{ma2024u} & 0.3451 ± 0.2568 & 62.7239 ± 20.1967 & 27.0104 ± 29.1163 & 0.2480 ± 0.2298 & 65.8801 ± 19.9775 & 22.5748 ± 15.4827 \\
U-KAN \cite{li2025u} & 0.5613 ± 0.2209 & 30.8672 ± 23.8262 & 9.8319 ± 13.3056 & 0.4750 ± 0.2591 & 34.2906 ± 25.1533 & 8.1040 ± 10.8974  \\
MDAN \cite{bao2021mdan} & 0.4703 ± 0.2304 & 44.7142 ± 21.7981 & 13.8409 ± 22.7259 & 0.3933 ± 0.2508 & 47.3511 ± 22.3255 & 8.9919 ± 12.2372 \\
SAN-Net \cite{yu2023san} & 0.5030 ± 0.2207 & 38.9887 ± 24.1235 & 16.2893 ± 33.1319 & 0.4412 ± 0.2505 & 41.0929 ± 24.8918 & 8.3931 ± 10.7925 \\
Cl-SegNet \cite{kuang2024hybrid} & 0.6154 ± 0.2101 & 29.2073 ± 24.4444 & 10.7944 ± 14.8114 & 0.5513 ± 0.2655 & 34.1256 ± 28.0769 & 8.7921 ± 11.2964 \\
\hline
\textbf{Our Method} & \textbf{0.6796 ± 0.1622*} & \textbf{23.5302 ± 21.7222*} & \textbf{7.6883 ± 13.0843*} & \textbf{0.6131 ± 0.2317*} & \textbf{26.0190 ± 24.1697*} & \textbf{5.9900 ± 10.7770*} \\
\hline
\end{tabular}
\end{adjustbox}
\label{tab:seg_results}
\end{table*}

\subsection{Computational Cost Analysis}
\label{subsec:computational_cost}
\begin{figure}[t]
\centering
\includegraphics[width=0.6\columnwidth]{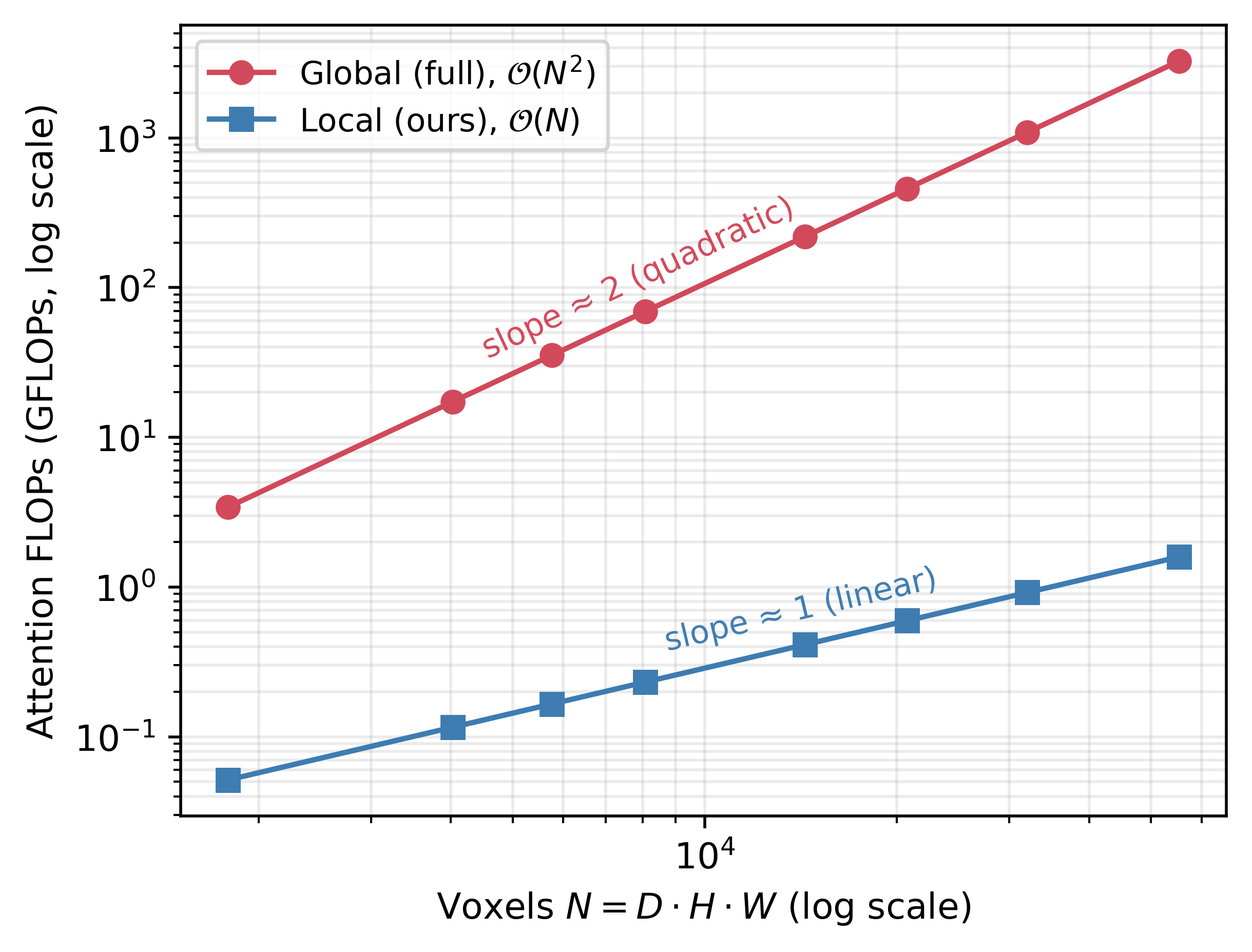}
\caption{Attention cost versus input size ($N=D\!\times\!H\!\times\!W$) for full attention (red) and CSA (blue). Full attention scales quadratically ($\mathcal{O}(N^2)$) while our CSA scales linearly ($\mathcal{O}(N)$).}
\label{fig:cost_scaling}
\end{figure}

To isolate the efficiency gain from restricting CSA to a local neighborhood, we compare it against an otherwise identical \emph{full} (global) cross-attention variant, differing only in attention scope, profiled at a representative resolution of $14\times32\times32$ ($N=14{,}336$), $C=256$ on a single NVIDIA RTX A6000 (48\,GB), batch size 1. As summarized in Table~\ref{tab:cost}, both variants share identical parameter counts, confirming the observed savings stem entirely from the attention mechanism rather than model capacity. Restricting attention to a $3\times3\times3$ neighborhood reduces attention-specific FLOPs by $531\times$ and total block FLOPs by $28.5\times$, cutting peak training memory from $24.6$\,GB to $1.6$\,GB ($15.6\times$) and inference time by $11.7\times$. Fig.~\ref{fig:cost_scaling} further shows that global attention scales quadratically ($\mathcal{O}(N^2)$) with input size, while the proposed local CSA scales linearly ($\mathcal{O}(N)$), consistent with the complexity argument in Section~\ref{heading:csa}. 

\begin{table}[t]
\centering
\caption{Cost of local (ours) vs.\ full cross-attention at
$14\times32\times32$, $C=256$ ($N=14{,}336$), batch size~1, single GPU.
Ratio = global\,/\,local. FLOPs use MAC~$=2$ operations.}
\label{tab:cost}
\footnotesize
\setlength{\tabcolsep}{4pt}
\renewcommand{\arraystretch}{1.0}
\begin{tabular}{lccc}
\toprule
\textbf{Metric} & \textbf{Local (Ours)} & \textbf{Global} & \textbf{Ratio} \\
\midrule
Parameters              & 263{,}168 & 263{,}168 & 1.0$\times$ \\
Total FLOPs (G)         & 7.93      & 226.19    & 28.5$\times$ \\
Attention FLOPs (G)        & 0.41      & 218.67    & 531$\times$ \\
Inference time (ms)     & 6.17      & 72.42     & 11.7$\times$ \\
Training time (ms)      & 27.39     & 227.27    & 8.3$\times$ \\
Inference peak memory (GB)   & 1.25      & 12.36     & 9.9$\times$ \\
Train peak memory (GB)     & 1.58      & 24.60     & 15.6$\times$ \\
\bottomrule
\end{tabular}
\end{table}

\subsection{Volumetric Analysis} 
\label{sec:vol_analysis}
\begin{figure}[ht]
\centering 
\includegraphics[width=0.8\columnwidth]{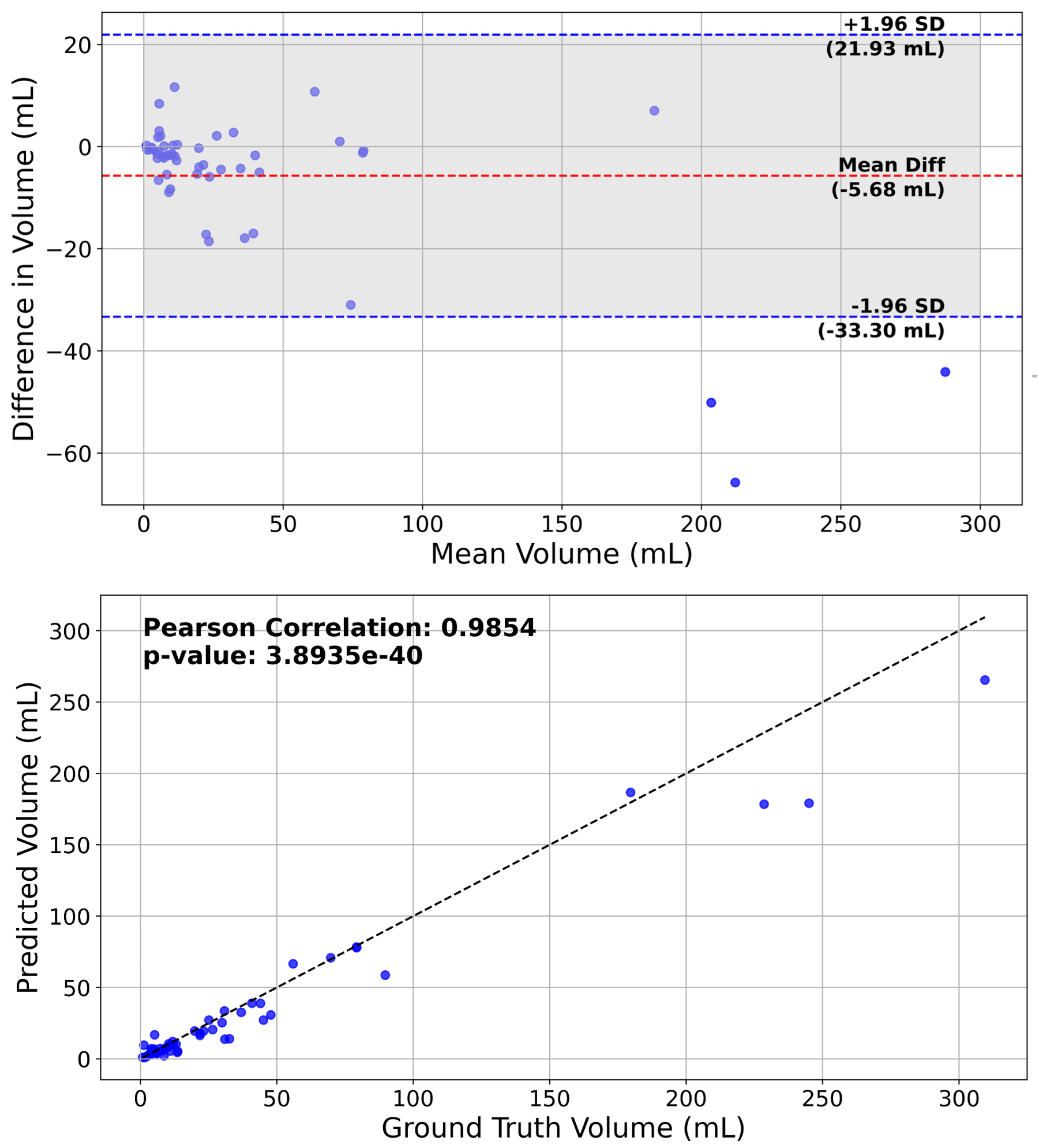} 
\caption{\textbf{Volumetric Analysis for AISD} (top) Bland-Altman plot comparing predicted volume with the ground truth, (bottom) Scatter plot of the predicted volumes versus the ground truth.}
\label{fig:vol_plots} 
\end{figure}

Lesion volume plays an important role in treatment planning for AIS, with a 70\,mL cutoff commonly used to determine patients' eligibility for intravenous thrombolysis and mechanical thrombectomy \cite{turc2023european}. We adopt this cutoff to stratify infarcts as small or large. As shown in Table~\ref{tab:seg_results}, our method achieves the highest mean Dice of 0.6131 outperforming every other state-of-the-art methods statistically at segmenting small lesions. Fig.~\ref{fig:vol_plots}(a) shows a Bland-Altman plot of predicted versus ground-truth lesion volume on AISD: the mean difference was $-5.68$\,mL, with limits of agreement from $-33.30$ to $21.93$\,mL, indicating good agreement with a bias close to zero and spread of differences remaining relatively contained within the limits of agreement. Fig.~\ref{fig:vol_plots}(b) shows the corresponding scatter plot, yielding a Pearson correlation of $r = 0.9854$ ($p < 0.001$), confirming a strong linear relationship between predicted and ground-truth volumes.

\subsection{Ablation Studies}
\label{subsec:ablation}
We conduct an extensive set of ablation studies on AISD to isolate the contribution of each architectural component; results are summarized in Table~\ref{tab:ablation}.
\subsubsection{Study I - Effectiveness of the overall framework} 
\label{sssec:framework}
We evaluate three configurations relative to the full method: (i) single-channel input, no tilt correction, no AsymFeX (plain 5-layer U-Net) (ii) dual-channel input (original and flipped) without tilt correction, retaining AsymFeX, isolating Stage~1's effect and (iii) dual-channel, tilt-corrected input with AsymFeX removed, isolating AsymFeX's architectural contribution. Configuration~(i) degrades Dice by 8.1\% and HD95 by 34.5\% relative to the full method, confirming the overall framework's effectiveness. Configuration~(ii) degrades Dice by 5.6\% and HD95 by 52.27\%, which is a sharper HD95 drop than removing AsymFeX alone, confirming tilt correction is a necessary prerequisite for exploiting the contralateral symmetry prior. Configuration~(iii) degrades Dice by 4.9\% and HD95 by 27.79\%. Notably, dual-channel input without AsymFeX (Table~\ref{tab:ablation}, row 3) improves only marginally over the single-channel baseline (Dice $0.6463$ vs.\ $0.6246$), while the full method recovers substantially more (Dice $0.6796$), indicating the gain stems primarily from the proposed comparison mechanism rather than additional input alone. For all configurations, the improvements are statistically significant ($p<0.05$).

\subsubsection{Study II - Placement of the AsymFeX modules} 
\label{sssec:asymfex}
We assess two configurations to assess AsymFeX's contribution at each encoder scale (layers 4 and 5): retaining AsymFeX only at layer 5 (removed from layer 4), and retaining it only at layer 4 (removed from layer 5). The former degrades Dice by 3.0\% and increases HD95 by 14.49\%; the latter degrades Dice by 2.2\% and increases HD95 by 9.7\%, indicating that asymmetric cues at both spatial scales contribute complementary information.

\subsubsection{Study III - Effectiveness of sub-modules under AsymFeX} 
\label{sssec:submodules}
We evaluate three configurations, each removing or replacing one AsymFeX sub-module while retaining the other two: (i) CSA removed entirely (ii) FDE replaced with naive element-wise feature subtraction and (iii) DSG removed entirely, with performance additionally assessed via the Absolute Lesion Count Difference (ALCD) to specifically test DSG's role in preserving small, discrete lesions. 
Removing CSA significantly degrades Dice by 6.32\% ($p<0.05$) and HD95 by 37.76\%, the largest single-component drop observed, underscoring the importance of localized cross-hemispheric attention. Replacing FDE with naive subtraction degrades Dice by 2.9\% and HD95 by 24.63\%. Removing DSG significantly degrades Dice by 3.7\% ($p<0.05$) and HD95 by 36.52\%. The overall mean ALCD also rises significantly ($p<0.05$) from $2.19$ to $3.48$ (a 37.0\% increase). Furthermore, applying the same 70\,mL threshold we evaluate the w/o DSG configuration on small lesions ($\le70$mL), and obtain a mean dice of 0.5877, mean HD95 of 34.02 and mean ALCD of 3.48. This is a significant degradation of performance in segmenting small lesions compared to our full configuration (Table \ref{tab:seg_results}) indicating DSG's particular importance for correctly identifying small, discrete lesions. 

We additionally conduct an additive ablation: (i) CSA alone, (ii) CSA+FDE (equivalent to w/o DSG), and (iii) the full CSA+FDE+DSG method. CSA alone degrades Dice by 5.02\% and HD95 by 9.7\%. Adding FDE improves Dice to $0.6544$ but temporarily raises HD95 to $32.12$, indicating that while FDE sharpens the disparity signal, the model responds with inflated boundary error without DSG to regulate it. Adding DSG resolves this, yielding the full model's Dice of $0.6796$ and HD95 of $23.53$. Both intermediate differences along this additive path are statistically significant, with the gap narrowing progressively as each module is added, confirming each component's contribution to AsymFeX's overall effectiveness.

\subsubsection{Study IV - Selection of Spatial Neighborhood}
\label{sssec:neighborhood}
To justify the $3\times3\times3$ CSA neighborhood, we evaluate two alternatives: a $1\times1\times1$ window, where each voxel in $Q$ attends only to its exact counterpart in $K$ with no neighborhood context, and an enlarged $5\times5\times5$ window, testing whether additional context is beneficial or introduces noise from distant tissue.
The $1\times1\times1$ window degrades Dice by 2.2\% and HD95 by 13.15\%, confirming that neighborhood context is essential to accommodate minor misalignments and anatomical shifts. The $5\times5\times5$ window degrades Dice by 3.1\% and HD95 by 13.70\%, likely from incorporating redundant information from distant, healthy tissue. Both differences in Dice values are significant ($p<0.05$) implying that $3\times3\times3$ configuration is thus both necessary and sufficient, indicating an efficiency-accuracy tradeoff point rather than an arbitrary choice.

\begin{table}
\centering
\footnotesize
\caption{Ablation study on AISD. Best result is highlighted in
\textbf{bold}.}
\setlength{\tabcolsep}{4pt}
\renewcommand{\arraystretch}{1.25}
\begin{adjustbox}{max width=\columnwidth}
\begin{tabularx}{\columnwidth}{>{\raggedright\arraybackslash}X|r|r}
\hline
\textbf{Ablation Experiment} & \multicolumn{1}{c|}{\textbf{Dice} $\uparrow$} & \multicolumn{1}{c}{\textbf{HD95} $\downarrow$} \\
\hline
Single-channel, w/o tilt correction, w/o AsymFex & 0.6246 & 31.6476 \\ 
Dual-channel, w/o tilt correction, with AsymFeX & 0.6415 & 35.8302 \\
Dual-channel, with tilt correction, w/o AsymFeX & 0.6463 & 30.0690 \\
\hline
AsymFeX only in layer 5 & 0.6590 & 26.9394 \\
AsymFeX only in layer 4 & 0.6648 & 25.8136 \\
\hline
w/o CSA & 0.6366 & 32.4149 \\
with CSA only & 0.6454 & 27.3981 \\
w/o FDE & 0.6602 & 29.3264 \\
with CSA + FDE (same as w/o DSG) & 0.6544 & 32.1235 \\
\hline
AsymFeX with neighborhood size $1^3$ & 0.6644 & 26.6024 \\
AsymFeX with neighborhood size $5^3$ & 0.6585 & 26.7541 \\
\hline
\textbf{Our Method} & \textbf{0.6796} & \textbf{23.5302} \\
\hline
\end{tabularx}
\end{adjustbox}
\label{tab:ablation}
\end{table}

\begin{figure}[!h]
\centering 
\includegraphics[width=0.85\columnwidth]{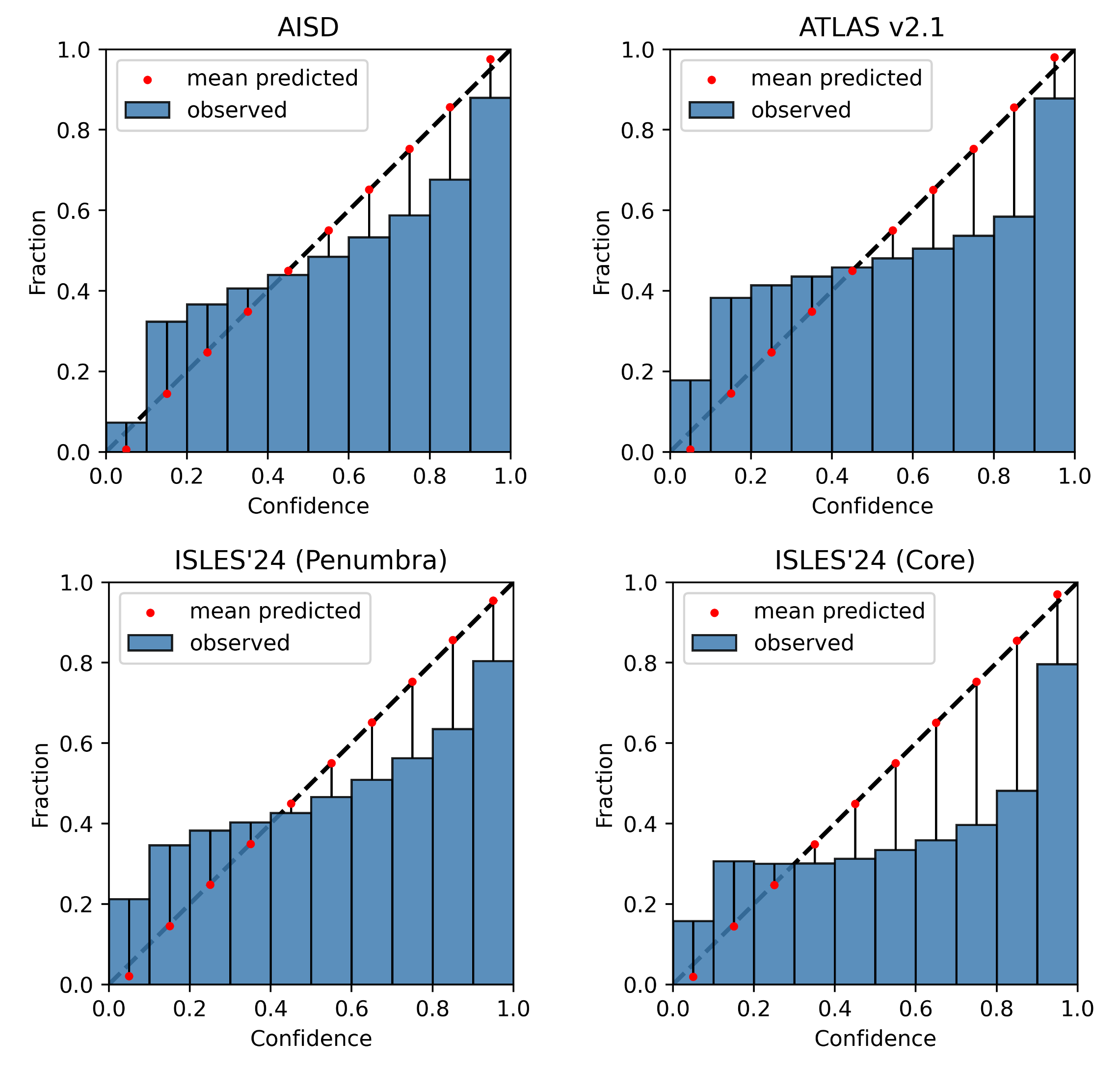}
\caption{\textbf{Calibration plots.} Bars show the observed foreground frequency per bin, red markers show the mean predicted probability and diagonal denotes perfect calibration. Deviations from the bars indicate miscalibration. The corresponding ECE values are reported in Table~\ref{tab:uncertainty_metrics}.}
\label{fig:cal_plots} 
\end{figure}

\subsection{Modality-Specific Adaptations and Cross-Dataset Comparison Protocol}
\label{subsec:mod_spfc_adap}
The core premise of the proposed framework is that loss of left-right hemispheric symmetry during an ischemic event comes from brain anatomy and is observable across imaging modalities and stroke timepoints, from acute hypoattenuation on NCCT, to hypoperfusion on CT perfusion, to chronic infarction on MRI. To assess whether the proposed architecture generalizes across this continuum across multiple modalities without modification, we evaluate it on ATLAS v2.1 \cite{liew2022large} (sub-acute and chronic, T1w MRI) and ISLES'24 \cite{de2024isles,riedel2024isles} (acute, CT perfusion), adapting only the Stage~1 preprocessing to each modality's input characteristics.

For ATLAS v2.1, a single-modality T1w dataset, the network receives the original volume and its flipped copy, identical to the primary AISD configuration. Since the dataset already had pre-processed T1-w MRI in MNI space, additional tilt-correction is not required. For ISLES'24, the tilt angle and translation are computed once from the NCCT channel and applied identically to the co-registered CBV, CBF, MTT, and $T_{\max}$ maps, rather than re-derived per channel, ensuring all channels remain spatially consistent after alignment. The encoder then receives all five channels (NCCT, CBV, CBF, MTT, $T_{\max}$) alongside their corresponding flipped copies. No further architectural modification is made in either case.

Having already reported a full ablation on AISD (Section~\ref{subsec:ablation}), we restrict this evaluation to a focused comparison against the five best-performing baseline methods from the AISD benchmark, together with two methods designed specifically for each target modality. As reported in Table~\ref{tab:isles_atlas_results}, the proposed method achieves the highest Dice on the infarct class on both ATLAS and ISLES'24, provides within 0.01 Dice of the best-performing method (nnU-Net) on the penumbra class, and attains the lowest HD95 across all compared methods on both datasets. On ATLAS, our method performs statistically better than all compared methods ($p < 0.05$). Also, on ISLES'24, without any additional architectural changes beyond the input-fusion strategy described above, our method performed on par or provided better results compared to existing methods. This superior performance relative to methods purpose-built for each modality, achieved without any modality-specific architectural tuning, instead supports the generality of the proposed asymmetry-driven design across the stroke imaging continuum, which is the central claim of this proof-of-concept evaluation.

\begin{table*}[t]
\centering
\caption{Quantitative comparison of the proposed method against existing
approaches on ATLAS v2.1 and ISLES'24. Symbol * denotes statistically significant
difference in performance with p-value $<$ 0.05 using Wilcoxon signed rank test.}
\label{tab:isles_atlas_results}
\begin{subtable}[t]{0.375\textwidth}
\centering
\caption{ATLAS v2.1}
\label{tab:atlas}
\begin{adjustbox}{max width=\linewidth}
\begin{tabular}{l|c|c}
\hline
\textbf{Method} & \makecell{\textbf{Dice} $\uparrow$ \\ (mean $\pm$ std.)} & \makecell{\textbf{HD95 (mm)} $\downarrow$ \\ (mean $\pm$ std.)} \\
\hline
nnUNet (3D)\cite{isensee2021nnu} & 0.6131 ± 0.2719 & 25.9931 ± 25.8474 \\
Attention U-Net \cite{oktay2018attention} & 0.6101 ± 0.2677 & 26.4353 ± 26.4768 \\
U-Net++ \cite{zhou2018unet++} & 0.6210 ± 0.2699 & 25.3857 ± 27.4043 \\
U-Mamba Bot \cite{ma2024u} & 0.6168 ± 0.2694 & 25.6742 ± 26.5214 \\
U-KAN \cite{li2025u} & 0.5857 ± 0.2885 & 24.0296 ± 25.7568 \\
MDAN \cite{bao2021mdan} & 0.4884 ± 0.2817 & 32.6047 ± 26.6804 \\
SAN-Net \cite{yu2023san} & 0.5073 ± 0.3114 & 33.6229 ± 30.6975 \\
\hline
\textbf{Our Method} & \textbf{0.6353 ± 0.2597*} & \textbf{20.9578 ± 23.4676*} \\
\hline
\end{tabular}
\end{adjustbox}
\end{subtable}
\hfill
\begin{subtable}[t]{0.605\textwidth}
\centering
\caption{ISLES'24 (Pre-treatment AIS segmentation)}
\label{tab:isles}
\begin{adjustbox}{max width=\linewidth}
\begin{tabular}{l|c|c|c|c}
\hline
\multirow{2}{*}{\textbf{Method}} & \multicolumn{2}{c|}{\textbf{Infarct}} & \multicolumn{2}{c}{\textbf{Penumbra}} \\
\cline{2-5}
 & \makecell{\textbf{Dice} $\uparrow$ \\ (mean $\pm$ std.)} & \makecell{\textbf{HD95 (mm)} $\downarrow$ \\ (mean $\pm$ std.)} & \makecell{\textbf{Dice} $\uparrow$ \\ (mean $\pm$ std.)} & \makecell{\textbf{HD95 (mm)} $\downarrow$ \\ (mean $\pm$ std.)} \\
\hline
nnUNet (3D)\cite{isensee2021nnu} & 0.5885 ± 0.2708 & 19.0980 ± 22.2467 & \textbf{0.6340 ± 0.2233} & 22.1314 ± 23.5837 \\
Attention U-Net \cite{oktay2018attention} & 0.5672 ± 0.2852 & 22.4399 ± 25.0435 & 0.6242 ± 0.2201 & 27.7271 ± 25.9984 \\
U-Net++ \cite{zhou2018unet++} & 0.5682 ± 0.2754 & 19.6201 ± 22.7191 & 0.6139 ± 0.2151 & 26.7082 ± 25.1919 \\
U-Mamba Bot \cite{ma2024u} & 0.6191 ± 0.2598 & 15.4238 ± 19.4771 & 0.6158 ± 0.2388 & 23.0358 ± 25.4424 \\
U-KAN \cite{li2025u} & 0.5686 ± 0.2811 & 17.1809 ± 16.9213 & 0.5889 ± 0.2280 & 20.1972 ± 19.2605 \\
ISP-Net \cite{ZHU2022106630} & 0.5230 ± 0.2801 & 20.4960 ± 21.2907 & 0.5889 ± 0.2087 & 23.0446 ± 22.1963 \\
I2PC-Net \cite{kuang2024segmenting} & 0.5901 ± 0.2771 & 16.2586 ± 19.3044 & 0.6209 ± 0.2469 & 22.9554 ± 23.0738 \\
\hline
\textbf{Our Method} & \textbf{0.6354 ± 0.2520} & \textbf{15.1846 ± 17.2990} & 0.6339 ± 0.2348 & \textbf{13.8642 ± 13.7755} \\
\hline
\end{tabular}
\end{adjustbox}
\end{subtable}
\end{table*}

\subsection{Perfusion Necessity on ISLES'24}
\label{ssec:perfu_necessity}
Given the short mean onset-to-door time on ISLES'24 (3h:28m \cite{riedel2024isles}), ischemic infarcts typically lack sufficient hypoattenuation to be visible on NCCT within this interval. To verify if NCCT alone is inadequate at this time point, we segment AIS lesions using NCCT only, keeping architecture and training protocol unchanged, for our method and the two best-performing methods from Table~\ref{tab:seg_results}. All three methods collapse to a near-identical, near-failing Dice on NCCT alone (nnU-Net (3D): $0.1766 \pm 0.1492$; Attention U-Net: $0.1763 \pm 0.1481$; Ours: $0.1786 \pm 0.1498$), with HD95 uniformly poor across all three ($159$-$163$\,mm), compared to a Dice of $0.6354$ once perfusion maps are included. This near-identical collapse across architectures indicates the limitation is intrinsic to NCCT's visibility of AIS lesions at this time point rather than any specific architectural shortcoming, confirming that CT perfusion parameter maps are necessary, not merely beneficial, for ISLES'24.

\subsection{Uncertainty Analysis}
\label{sec:uncert}
To evaluate the reliability of the model's predictions, we quantify predictive uncertainty using Monte Carlo (MC) dropout at inference. Dropout layers (rate $10\%$) placed after the final AsymFeX block in the last two encoder layers remain active during testing, and each test case is passed through 30 stochastic forward passes. We validate the resulting uncertainty estimates via two complementary analyses: an Error-Uncertainty analysis, testing whether high-uncertainty regions coincide with actual segmentation errors, and a Calibration analysis, testing whether predicted confidence matches observed correctness.

\subsubsection{Error-Uncertainty Analysis}
\label{subsubsec:error_uncert_analysis}
A reliable model should exhibit high uncertainty precisely where it errs. We evaluate this within a region of interest (ROI) defined as the union of the predicted and ground-truth masks, dilated slightly to exclude trivial background voxels. Within this ROI, each voxel's uncertainty is treated as a predictor of segmentation error, and we report the Area Under the ROC Curve (AUROC) and Area Under the Precision-Recall Curve (AUPRC) against the corresponding error prevalence. We also report mean uncertainty stratified by True Positive (TP), False Positive (FP), False Negative (FN), and True Negative (TN) voxels. As shown in Table~\ref{tab:uncertainty_metrics}, AUROC exceeds $0.70$ and AUPRC exceeds the corresponding prevalence for every dataset, confirming that predicted uncertainty identifies erroneous voxels well above chance. Moreover, mean uncertainty is consistently higher for error categories (FP, FN) than for correct categories (TP, TN) across all datasets.

\subsubsection{Calibration Analysis.} 
\label{sssec:calibration}
To assess whether predicted probabilities match observed correctness, Fig.~\ref{fig:cal_plots} shows reliability diagrams for each dataset (observed frequency vs.\ the diagonal line of perfect calibration), with the corresponding Expected Calibration Error (ECE) reported in Table~\ref{tab:uncertainty_metrics}; both are computed over the ROI defined in Section~\ref{subsubsec:error_uncert_analysis}. AISD shows the strongest calibration among the three datasets, while ATLAS and ISLES'24 are moderately miscalibrated. The higher ECE on ISLES'24 is consistent with genuine ambiguity at the infarct-penumbra boundary, which is typically diffuse rather than sharp, producing many inherently ambiguous voxels; this is further supported by the higher mean uncertainty of correctly classified voxels on ISLES'24 relative to AISD and ATLAS (Table~\ref{tab:uncertainty_metrics}). ISLES'24's smaller training set (109 cases) could also have contributed to this miscalibration.

Taken together with the Dice results in Sections~\ref{subsec:mod_spfc_adap} and \ref{ssec:perfu_necessity}, AISD achieves both the highest accuracy and the sharpest boundary confidence among the three datasets, while ISLES'24 attains comparable Dice but markedly higher boundary uncertainty. This indicates that the method is not merely equally accurate across the stroke continuum, but less \emph{confidently} so at earlier, perfusion-dependent time points where the underlying tissue distinction is itself more ambiguous. 
\begin{table}[t]
\centering
\caption{Uncertainty metrics per dataset. All values are computed within the foreground ROI. Every metric is calculated over the MC-averaged
T=30 passes probabilities.}
\label{tab:uncertainty_metrics}
\footnotesize
\renewcommand{\arraystretch}{1.15}
\begin{tabular*}{\columnwidth}{@{\extracolsep{\fill}} l cc cc @{}}
\toprule
\multirow{2}{*}{\textbf{Metric}} & \multirow{2}{*}{\textbf{AISD}}
 & \multirow{2}{*}{\textbf{ATLAS v2.1}}
 & \multicolumn{2}{c}{\textbf{ISLES'24}} \\
\cmidrule(lr){4-5}
 & & & \textbf{Infarct} & \textbf{Penumbra} \\
\midrule
AUROC          & 0.8102 & 0.7277 & 0.7706 & 0.7053 \\
AUPRC          & 0.3678 & 0.3832 & 0.4819 & 0.4357 \\
Prevalence  & 0.1296 & 0.1947 & 0.2515 & 0.2837 \\
\midrule
\multicolumn{5}{@{}l}{\textit{Mean uncertainty}}\\
\quad TP       & 0.1719 & 0.2206 & 0.3028 & 0.4545 \\
\quad TN       & 0.0739 & 0.0896 & 0.2479 & 0.3070 \\
\quad FP       & 0.4999 & 0.4776 & 0.6566 & 0.6624 \\
\quad FN       & 0.2880 & 0.2425 & 0.4825 & 0.5498 \\
\midrule
ECE            & 0.0807 & 0.1437 & 0.1694 & 0.1660 \\
\bottomrule
\end{tabular*}
\end{table}
\section{Discussion and Conclusion}
\label{sec:disc_conc}
In this study, we propose a two-stage, 3D AIS segmentation pipeline that incorporates a stroke-specific prior and integrates directly with nnU-Net. At its core is AsymFeX, which combines local 3D cross-hemispheric attention, disparity estimation, and dual-scale gating, restricting comparison to a local neighborhood around each voxel's true contralateral landmark rather than the entire opposite hemisphere, closely mirroring how a radiologist compares corresponding regions across hemispheres in practice. Across an extensive evaluation on AISD, our method achieves the best Dice, HD95, and AVD among nine baselines and three stroke-specific state-of-the-art methods, with gains statistically significant via Wilcoxon signed-rank testing (Table \ref{tab:seg_results}). Ablation confirms that both stages of the pipeline are necessary, bypassing tilt correction alone degrades HD95 more sharply than removing AsymFeX entirely, and dual-channel tilt-corrected input without AsymFeX yields only marginal gains over a single-channel baseline, confirming that the observed improvement stems from the proposed comparison mechanism rather than additional input alone (Table \ref{tab:ablation}). Volumetric analysis shows strong agreement with ground-truth lesion volume (Pearson $r = 0.9854$, Fig. \ref{fig:vol_plots}), directly relevant to the 70\,mL threshold used clinically to determine thrombolysis and thrombectomy eligibility. 

Proof-of-concept evaluation on ATLAS v2.1 and ISLES'24 shows the same architecture, adapted only at the input level, remains competitive with modality-specialized methods without any architectural retuning. The perfusion-necessity experiment on ISLES'24 confirms that this adaptation is not incidental: NCCT alone collapses to near-failing performance at this timepoint regardless of architecture, underscoring that imaging modality, not model design, is the binding constraint at early stroke stages. Uncertainty analysis further shows that boundary confidence and accuracy varies across the stroke continuum. AISD yields both the highest accuracy and the sharpest boundary confidence (Table \ref{tab:uncertainty_metrics}), while ISLES'24 achieves comparable Dice with markedly higher boundary uncertainty, indicating the model is \emph{not equally confident} across all time points even where it is equally accurate. Collectively, these results carry clear implications for clinical translation. The strong volumetric agreement near the 70\,mL thrombolysis/thrombectomy eligibility threshold \cite{seners2023quantification} suggests the method could support rapid, automated volume estimation at triage, where treatment delay directly reduces salvageable penumbra \cite{seners2023quantification}\cite{saver2006time}. Generalization across NCCT, CT perfusion, and MRI without architectural redesign is relevant given that multiple modalities could be acquired for a given patient depending on onset time or to gather all the necessary information for effective treatment planning \cite{heit2015imaging}, favoring a single adaptable pipeline over separate per-modality models. The uncertainty framework further offers calibrated clinical trust rather than a bare prediction, particularly relevant given that boundary confidence degrades at earlier, more treatment-critical stroke stages. 

The proposed framework assumes the two hemispheres are approximately mirror-symmetric prior to infarction, an assumption more likely violated by substantial midline shift or, in chronic cases, encephalomalacia and ventricular enlargement unrelated to the acute event, both of which may compromise Stage~1 alignment and downstream AsymFeX comparison. Also, the dual-stream encoder also incurs meaningfully greater inference cost than a single-stream baseline despite the efficiency of windowed attention (Section~\ref{subsec:computational_cost}), which may matter in resource-constrained settings. 
Three future research directions follow naturally. Using uncertainty estimates potentially from reporting to an operational role, flagging low-confidence cases for review, indicating when an additional modality is necessary, or guiding uncertainty-aware continual learning. Given the cost of manual delineation, training under weak supervision, using bounding boxes, image-level labels, or point annotations, could reduce annotation burden while retaining the asymmetry-driven comparison principle. Finally, given AIS's time-critical nature, extending the framework to predict lesion trajectory, i.e., the rate of penumbra-to-core conversion via collateral circulation information, would connect segmentation output more directly to treatment-window decisions.

\subsection*{Conclusions}
\label{ssec:conc}
In summary, this study demonstrates that an anatomically-grounded, computationally efficient asymmetry prior can deliver state-of-the-art AIS segmentation, improving Dice by 10.42\% over the strongest baseline and reducing HD95 by 19.44\% on AISD, while generalizing, without architectural redesign, across the acute-to-chronic stroke imaging continuum with strong volumetric agreement (Pearson $r = 0.9854$) and a $531\times$ reduction in attention FLOPs relative to global cross-attention, alongside reliability characteristics that vary in clinically interpretable ways across that continuum. Future directions include uncertainty-aware diagnosis, training under weak supervision and lesion trajectory prediction.

\bibliographystyle{IEEEtran}
\bibliography{refs}
\end{document}